\magnification \magstep1
\raggedbottom
\openup 2\jot
\voffset6truemm
\centerline {\bf A NEW APPLICATION OF NON-CANONICAL MAPS}
\centerline {\bf IN QUANTUM MECHANICS}
\vskip 1cm
\leftline {Giampiero Esposito}
\vskip 1cm
\noindent
{\it Istituto Nazionale di Fisica Nucleare, Sezione
di Napoli, Complesso Universitario di Monte S. Angelo, Via Cintia,
Edificio N', 80126 Napoli, Italy}
\vskip 0.3cm
\noindent
{\it Universit\`a di Napoli Federico II, Dipartimento
di Scienze Fisiche, Complesso Universitario di Monte S. Angelo,
Via Cintia, Edificio N', 80126 Napoli, Italy}
\vskip 1cm
\noindent
{\bf Abstract}. A proof is given that an invertible and a unitary operator 
can be used to reproduce the effect of a $q$-deformed 
commutator of annihilation and creation operators. 
In other words, the original annihilation and creation operators are
mapped into new operators, not conjugate to each other, 
whose standard commutator equals the 
identity plus a correction proportional to the original number operator.
The consistency condition for the existence of this new set of operators
is derived, by exploiting the Stone theorem on 1-parameter unitary groups.
The above scheme leads to modified
`equations of motion' which do not preserve the properties of the 
original first-order set for annihilation and creation operators. Their
relation with commutation relations is also studied.
\vskip 100cm
\leftline {\bf 1. INTRODUCTION}
\vskip 0.3cm
\noindent
Several efforts have been devoted in the literature to the attempt of
building quantum mechanics as a kind of deformed classical mechanics.
The mathematical foundations and the physical applications of such a
program are well described, for example in Ref. [1] and in the many
references given therein. Within that framework, quantization emerges
as an autonomous theory based on a deformation of the composition law
of classical observables, not on a radical change in the nature of the
observables. One then gets a more general approach which coincides
with the conventional operatorial approach in known applications
whenever a Weyl map can be defined, and leads to an improved 
conventional quantization in field theory [1]. 

In particular, this has led to consider the so-called $q$-deformed 
commutator of annihilation and creation operators of an harmonic
oscillator, i.e. [2]
$$
[a,a^{\dagger}]_{q} \equiv aa^{\dagger}-qa^{\dagger}a=I,
\eqno (1.1)
$$
$I$ being the identity operator.
It is the aim of this paper of providing an alternative interpretation
of Eq. (1.1) and discussing its implications, putting instead the emphasis
on maps which do not preserve the canonical commutation relations.
In other words, since non-canonical maps are an important topic in
quantum mechanics, we propose to exploit their properties to avoid having
to deform the composition law of observables. The following sections show
under which conditions this is indeed possible, and some of 
its implications.
\vskip 0.3cm
\leftline {\bf 2. A NEW LOOK AT DEFORMED COMMUTATORS}
\vskip 0.3cm
\noindent
For this purpose, we first point out that Eq. (1.1) can be re-expressed
in the form
$$
aa^{\dagger}-a^{\dagger}a=I+(q-1)a^{\dagger}a.
\eqno (2.1)
$$
Now the left-hand side of Eq. (2.1) is the application to $a$ and
$a^{\dagger}$ of the standard definition of commutator of a pair or
linear operators $A$ and $B$:
$$
[A,B] \equiv AB-BA,
\eqno (2.2)
$$
where, at this stage, we are leaving aside the technical problems 
resulting from the possible occurrence of unbounded operators [3,4].

The picture we have in mind is therefore as follows. Suppose we start 
from the operators $a$ and $a^{\dagger}$ satisfying the canonical 
commutation relations
$$
[a,a^{\dagger}]=I.
\eqno (2.3)
$$
Can we map $a$ and $a^{\dagger}$ into new operators $A$ and $B$ 
whose standard commutator satisfies instead the condition suggested by
Eq. (2.1), i.e.
$$
[A,B]=I+(q-1)N,
\eqno (2.4)
$$
having defined, as usual, $N \equiv a^{\dagger}a$ (the standard number
operator)? In other words, after re-writing Eq. (1.1) in the equivalent
form (2.1), we reinterpret {\it the left-hand side only} as the standard
commutator of new operators, here denoted by $A$ and $B$. By
doing so, we are aiming to prove that the standard commutator structure
in quantum mechanics can be preserved, while the mathematics of 1-parameter
unitary groups makes it possible to achieve a transition from Eq. (2.3)
to Eq. (2.4) (see also comments in section 6).

For this purpose, we consider a pair of invertible operators $S$
and $T$ chosen in such a way that $T$ is unitary
and the original commutation relation is no
longer preserved. This means that we define
$$
A \equiv SaT^{-1},
\eqno (2.5)
$$
$$
B \equiv T a^{\dagger} S^{-1},
\eqno (2.6)
$$
which implies that
$$
[A,B]=Saa^{\dagger}S^{-1}-Ta^{\dagger}aT^{-1},
\eqno (2.7)
$$
and eventually, from Eq. (2.3) and the definition of $N$,
$$
[A,B]=I+SNS^{-1}-TNT^{-1}.
\eqno (2.8)
$$
Note that $B$ is not even the formal adjoint of $A$, since $S$ is not
required to be unitary (which will be shown to be sufficient to realize
our non-canonical map).
Since we require that the commutator (2.8) should coincide with the commutator
(2.4), we obtain the equation
$$
SNS^{-1}=TNT^{-1}+(q-1)N.
\eqno (2.9)
$$

As we said already in section 1, we are dealing with maps which do not preserve
the canonical commutation relations. The non-linear map
$$
a \rightarrow \sqrt{[n]\over n} \; a
$$
provides an example of such a transformation. Our commutation relations (2.4)
are not the same as those of (1.1), for which
$$
[n]={{q^{n}-1}\over (q-1)}
$$
but correspond instead to 
$$
[n]=n+(q-1){n(n-1)\over 2}
$$
which is a polynomial deformation.
\vskip 0.3cm
\leftline {\bf 3. APPLICATION OF THE STONE THEOREM}
\vskip 0.3cm
\noindent
Having obtained the fundamental equation (2.9) we point out that, since
$T$ is taken to be unitary, we can exploit the Stone theorem [5], 
according to which to every weakly continuous, 1-parameter family
$U(s), s \in {\bf R}$ of unitary operators on a Hilbert space $\cal H$,
obeying 
$$
U(s_{1}+s_{2})=U(s_{1})U(s_{2}), \; \;
s_{1},s_{2} \in {\bf R},
\eqno (3.1)
$$
there corresponds a unique self-adjoint operator $A$ such that [3,4]
$$
U(s)={\rm e}^{isA},
\eqno (3.2)
$$
for all $s \in {\bf R}$. More precisely, the Stone theorem states that,
if $U(s)$, $s \in (-\infty, \infty)$, is a family of unitary
transformations with the group property (3.1) and such that 
$(U(s)f,g)$ is a measurable function of $s$ for arbitrary $f$ and $g$
in an abstract Hilbert space, then there exists a unique self-adjoint
operator $A$ such that $U(s)={\rm e}^{isA}$.

In our problem, we therefore consider a real parameter $u$ and a
self-adjoint operator $B$ such that
$$
T=T(u)={\rm e}^{iuB} \; \; u \in {\bf R}.
\eqno (3.3)
$$
We exploit Eq. (3.3) after choosing 
$B=P$ for convenience (see comments below),
i.e. the momentum operator canonically conjugate to the position
operator $Q$. 
In ${\hbar}=1$ units, the annihilation and creation operators read
$$
a \equiv {1\over \sqrt{2}}(Q+iP),
\eqno (3.4)
$$
$$
a^{\dagger} \equiv {1\over \sqrt{2}}(Q-iP),
\eqno (3.5)
$$
and hence the number operator can be written in the form
$$
N \equiv a^{\dagger}a={1\over 2}(Q^{2}+P^{2}-I).
\eqno (3.6)
$$
If
$$
T(u) \equiv {\rm e}^{iuP},
\eqno (3.7)
$$
we can exploit the identities
$$
{\rm e}^{-iuP} \; Q \; {\rm e}^{iuP}=Q-uI,
\eqno (3.8)
$$
$$
{\rm e}^{iuP} \; P=P \; {\rm e}^{iuP},
\eqno (3.9)
$$
to obtain
$$
TQT^{-1}=Q+uI,
\eqno (3.10)
$$
$$
TP=PT,
\eqno (3.11)
$$
and hence
$$ \eqalignno{
\; & TNT^{-1}={1\over 2}TQT^{-1}TQT^{-1}+{1\over 2}TPT^{-1}TPT^{-1}
-{I\over 2} \cr
&={1\over 2}(Q+uI)^{2}+{1\over 2}P^{2}-{I\over 2}
=N+uQ+{u^{2}\over 2}I.
&(3.12)\cr}
$$
It is now clear that the choice $B=P$ in (3.3), although not mandatory, 
is a matter of convenience, since it makes it possible to obtain a
manageable expression for $TNT^{-1}$.
This formula, resulting from the particular choice (3.7), can be inserted
into Eq. (2.9) which now becomes an equation for the unknown 
operator $S$, i.e.
$$
SNS^{-1}=qN+uQ+{u^{2}\over 2}I,
\eqno (3.13a)
$$
or also, more conveniently,
$$
S(Q^{2}+P^{2})S^{-1}=q(Q^{2}+P^{2})+2uQ
+(u^{2}-(q-1))I.
\eqno (3.13b)
$$
Now we consider the complete orthonormal set of harmonic oscillator states,
denoted by $| n \rangle$ with the abstract Dirac notation. On acting on both
sides of (3.13b) with $S$ from the right one finds
$$
S(2N+I)=q(2N+I)S+2uQS+(u^{2}-(q-1))S. 
\eqno (3.14)
$$
Since the task of finding $S$ is equivalent to the evaluation of all its
matrix elements, we point out that this equation leads to an equation for
matrix elements of $S$ upon exploiting the resolution of the identity
$$
I=\sum_{n=0}^{\infty}|n \rangle \langle n | ,
\eqno (3.15)
$$
when we write $S=SI$, and defining
$$
S_{m,n} \equiv \langle m | S | n \rangle.
\eqno (3.16)
$$
Since $N | m \rangle =m |m \rangle$, while $Q={1\over \sqrt{2}}
(a+a^{\dagger})$, one then finds, after evaluation of the bra 
$\langle m |$ on both sides of Eq. (3.14), the equation
$$ \eqalignno{
\; & \Bigr[((2n+1)-q(2m+1)-(u^{2}-(q-1)))S_{m,n} \cr
&-u\sqrt{2}(\sqrt{m}S_{m-1,n}+\sqrt{m+1}S_{m+1,n})\Bigr]=0,
&(3.17)\cr}
$$
where we have used the standard properties $a |m \rangle
=\sqrt{m} |m-1 \rangle, a^{\dagger} |m \rangle =\sqrt{m+1}
|m+1 \rangle$. Equation (3.17) implies that
$$
(2(n-mq)-u^{2})S_{m,n}=u\sqrt{2}(\sqrt{m}S_{m-1,n}
+\sqrt{m+1}S_{m+1,n}).
\eqno (3.18)
$$
For given values of $q$ and $u$, this set of equations should be studied
for all values of $n,m=0,1,...,\infty$. If $mq+{u^{2}\over 2}$ is not an
integer, this infinite set yields the matrix element $S_{m,n}$ as a linear
combination of $S_{m-1,n}$ and $S_{m+1,n}$, i.e.
$$
S_{m,n}=A_{mn}S_{m-1,n}+B_{mn}S_{m+1,n},
\eqno (3.19)
$$
where
$$
A_{mn}={u \sqrt{2m} \over (2(n-mq)-u^{2})} \; \; \;
B_{mn}={u \sqrt{2(m+1)}\over (2(n-mq)-u^{2})}.
\eqno (3.20)
$$
In agreement with our assumptions, these equations show that the
operator $S$ is not unitary, since it fails to satisfy the
basic condition $SS^{\dagger}=I$.

To prove the possibility of realizing $S$ as an invertible operator, we
find it more convenient to revert to the operator equation (3.14), here
written in the form
$$
S(Q^{2}+P^{2})=q(Q^{2}+P^{2})S+2uQS+\beta S ,
\eqno (3.21)
$$
having defined $\beta \equiv u^{2}-(q-1)$. Here the left- and
right-hand sides are operators acting on square-integrable stationary
states $\psi(x)$. In one spatial dimension, $Q$ can be realized as the
operator of multiplication by $x$, and $P$ as the operator 
$-i{d\over dx}$, if the coordinate representation is chosen. If $S$ is
taken to be the operator of multiplication by an 
invertible function $f$, i.e.
$$
S: \psi \rightarrow f(x)\psi(x)
$$
with $f: x \rightarrow f(x)$ invertible, we have to check that the
resulting differential equation for $\psi(x)$ admits square-integrable
solutions. Indeed, the choices outlined imply that Eq. (3.21) leads to
the following differential equation for $\psi(x)$:
$$
\left[{d^{2}\over dx^{2}}+\varphi_{1}(x){d\over dx}
+\varphi_{2}(x)\right]\psi(x)=0,
\eqno (3.22)
$$
where
$$
\varphi_{1}(x) \equiv {2q\over (q-1)}{f' \over f},
\eqno (3.23)
$$
$$
\varphi_{2}(x) \equiv -x^{2}+{-qf''+2uxf+\beta f \over (1-q)f}.
\eqno (3.24)
$$
To ensure that the origin is a regular singular point of Eq. (3.22) we
have to choose $f$ in such a way that $\varphi_{1}$ has, at most,
a first-order pole at $x=0$, and $\varphi_{2}$ has at most a second-order
pole at $x=0$. For example, such conditions are fulfilled if 
$f: x \rightarrow x$, because then $\varphi_{1}$ has a first-order pole
at $0$, while $\varphi_{2}$ has no poles at all therein, being equal to
$$
-x^{2}+{2ux + \beta \over (1-q)}.
$$
The resulting equation reads
$$
\left[{d^{2}\over dx^{2}}+{2q \over (q-1)}{1\over x}{d\over dx}
+\left(-x^{2}+{2ux +\beta \over (1-q)}\right)\right]\psi(x)=0,
\eqno (3.25)
$$
and for it the point at infinity is not Fuchsian, as it happens for the
ordinary harmonic oscillator in quantum mechanics. 

To sum up, we have shown that Eq. (3.21) is compatible with at least one
choice of invertible operator $S$ for which the stationary states are
square-integrable on the whole real line (the potential term in Eq. (3.25)
being dominated at large $x$ by an even function which diverges at
infinity). We have not considered the exponential map as a candidate for 
$f$ since its inverse, the logarithm, is not defined for negative $x$, while
the ordinary oscillator is studied for all values of $x$.
\vskip 0.3cm
\leftline {\bf 4. MODIFIED EQUATIONS OF MOTION}
\vskip 0.3cm
\noindent
In the investigation of deformed harmonic oscillators it is rather 
important to check that the equations of motion satisfied by the
annihilation and creation operators defined in (3.4) and (3.5), i.e.
$$
\left({d\over dt}+i \right)a=0,
\eqno (4.1)
$$
$$
\left({d\over dt}-i \right)a^{\dagger}=0,
\eqno (4.2)
$$
are preserved [6]. Here, however, we have mapped $(a,a^{\dagger})$ into
operators $(A,B)$ whose standard commutator satisfies
instead Eq. (2.4). It is therefore not obvious that the equations of
motion (4.1) and (4.2) are preserved. Indeed, 
by allowing for a time dependence of $T$ and $S$
one finds, by virtue of (2.5) and (4.1), that
$$
{dA\over dt}={\dot S}aT^{\dagger}+S{\dot a}T^{\dagger}
+Sa{{\dot T}}^{\dagger}
={\dot S}aT^{\dagger}+S(a{{\dot T}}^{\dagger}-iaT^{\dagger}).
\eqno (4.3)
$$
This leads to
$$
\left({d\over dt}+i \right)A={\dot S}aT^{\dagger}+Sa{{\dot T}}^{\dagger}.
\eqno (4.4)
$$
Now we would like to re-express the right-hand side of Eq. (4.4) in 
such a way that $a$ is replaced by $A$. For this purpose, we use 
Eq. (2.5), the unitarity of $T$ and the invertibility of $S$ to find
$$
aT^{\dagger}=S^{-1}A,
\eqno (4.5)
$$
$$
Sa=AT,
\eqno (4.6)
$$
and hence the operator $A$ obeys the first-order equation
$$
\left({d\over dt}+i \right)A={\dot S}S^{-1}A
+AT{{\dot T}}^{\dagger}.
\eqno (4.7)
$$
An analogous procedure shows that
$$
{dB\over dt}={\dot T}a^{\dagger}S^{-1}
+T \left(ia^{\dagger}S^{-1}+a^{\dagger}{dS^{-1}\over dt}\right),
\eqno (4.8)
$$
and hence
$$ 
\left({d\over dt}-i \right)B={\dot T}a^{\dagger}S^{-1}
+Ta^{\dagger}{dS^{-1}\over dt}
=BS{dS^{-1}\over dt}+{\dot T}T^{\dagger}B,
\eqno (4.9)
$$
where we have used the identities
$$
Ta^{\dagger}=BS ,
\eqno (4.10)
$$
$$
a^{\dagger}S^{-1}=T^{\dagger}B.
\eqno (4.11)
$$
\vskip 0.3cm
\leftline {\bf 5. EQUATIONS OF MOTION VS. COMMUTATORS}
\vskip 0.3cm
\noindent
In ordinary quantum mechanics one knows, thanks to the work of
Wigner [7] and other authors [8], that the equations of motion do not
determine uniquely the commutation relations one relies upon. In our
case, this amounts to asking whether, reversing the previous logical
order, Eqs. (4.7) and (4.9)
are more fundamental than the commutator (2.4), and to which extent a
solution of Eqs. (4.7) and (4.9) determines uniquely the commutator
of $A$ with $B$.

Indeed, on defining the first-order operators $\varphi \equiv {d\over dt}+i$ 
and $\gamma \equiv {d\over dt}-i$, and considering the commutators
$$
\Bigr[A,T{{\dot T}}^{\dagger}\Bigr] \equiv C_{1},
\eqno (5.2)
$$
$$
\Bigr[B,S {dS^{-1}\over dt}\Bigr] \equiv C_{2},
\eqno (5.3)
$$
Eqs. (4.7) and (4.9) can be written in the form
$$
\Bigr(\varphi-{\dot S}S^{-1}-T{{\dot T}}^{\dagger}\Bigr)A
=C_{1},
\eqno (5.4)
$$
$$
\Bigr(\gamma-S{dS^{-1}\over dt}-{\dot T}T^{\dagger}\Bigr)B 
=C_{2}.
\eqno (5.5)
$$
The resulting analysis, far from being of purely formal value, goes
at the very heart of the problem: one can solve for $A$ and
$B$ upon inverting the operators in round brackets in Eqs.
(5.4) and (5.5), and this makes it necessary to find their Green
functions. But there may be more than one Green function, depending
on which initial condition is chosen. Assuming that such a choice
has been made, one can write
$$
A=\Bigr(\varphi-{\dot S}S^{-1}-T{{\dot T}}^{\dagger}\Bigr)^{-1}C_{1},
\eqno (5.6)
$$
$$
B=\Bigr(\gamma-S{dS^{-1}\over dt}-{\dot T}T^{\dagger}
\Bigr)^{-1}C_{2},
\eqno (5.7)
$$
and their commutator is not obviously equal to (see (2.4))
$$
I+(q-1)a^{\dagger}a=I+(q-1)T^{\dagger}BAT
$$
where we have inverted the equations (2.5) and (2.6) defining $A$ and
$B$ to find
$$
a=S^{-1}AT, \; a^{\dagger}=T^{\dagger}BS.
$$
\vskip 0.3cm
\leftline {\bf 6. CONCLUDING REMARKS}
\vskip 0.3cm
\noindent
Starting from Eqs. (1.1) and (2.1) we have pointed out that deformed
commutators can be ``replaced'' by a map of the standard commutation
relations (2.3) into the modified form (2.4). 
As far as we can see, this is by no means
equivalent to deformation quantization. Our effort to build such a map
reflects instead the desire to preserve the standard commutator
structure, while using some basic mathematical tools to prove
that the map of Eq. (2.3) into Eq. (2.4) is feasible. This leads to
the introduction of two different invertible operators $S$ and $T$
with $T$ unitary, 
subject to the consistency condition (2.9). From the point of view
of ideas and techniques, this is the original contribution of our paper.
Section 3 proves that a
careful use of the Stone theorem makes it possible to fulfill such a
condition with $S$ invertible,  
while sections 4 and 5 have studied how the equations of
motion are modified, and what sort of correspondence exists between
them and the commutator (2.4). 

Our framework can be made broader by studying the case when neither
$S$ nor $T$ is unitary (see (2.5) and (2.6),
but we see no (obvious) advantage in doing so. 
Our investigation is of interest for the mathematical 
foundations of quantum mechanics, because {\it it shows under which
conditions it is possible to avoid deforming the composition
law of classical observables} (cf. Refs. [1,9--13]). 
Further developments can also be expected, because the link between the
superoperator formalism [14] and the maps defined by our equations
(2.5) and (2.6) deserves a thorough investigation.
\vskip 0.3cm
\leftline {\bf ACKNOWLEDGMENTS}
\vskip 0.3cm
\noindent
The work of G. Esposito has been partially supported by
the Progetto di Ricerca di Interesse Nazionale
{\sl SINTESI 2000}. He is grateful to
Volodia Man'ko and Stefan Weigert for detailed correspondence.
\vskip 0.3cm
\leftline {\bf REFERENCES}
\vskip 0.3cm
\item {[1]}
D. Sternheimer (1998). {\it Deformation Quantization: Twenty Years Later}
(math.QA 9809056).
\item {[2]}
M. Arik and D. D. Coon (1976). {\it J. Math. Phys.} {\bf 17}, 524.
\item {[3]}
M. Reed and B. Simon (1972). {\it Methods of Modern Mathematical Physics.
Vol. I: Functional Analysis} (Academic Press, New York). 
\item {[4]}
E. Prugovecki (1981). {\it Quantum Mechanics in Hilbert Space. II Edition}
(Academic Press, New York). 
\item {[5]} 
M. H. Stone (1932). {\it Ann. Math.} {\bf 33}, 643. 
\item {[6]} 
V. I. Man'ko, G. Marmo and F. Zaccaria (1996). 
{\it Rend. Sem. Mat. Univ. Pol. Torino} {\bf 54}, 337. 
\item {[7]}
E. P. Wigner (1950). {\it Phys. Rev.} {\bf 77}, 711. 
\item {[8]}
V. I. Man'ko, G. Marmo, E. C. G. Sudarshan and F. Zaccaria (1996). 
{\it Wigner's Problem and Alternative Commutation Relations for 
Quantum Mechanics} (QUANT-PH 9612007).
\item {[9]}
F. Bayen, M. Flato, C. Fronsdal, A. Lichnerowicz and
D. Sternheimer (1978). {\it Ann. Phys. (N.Y.)} {\bf 111}, 61;
{\it Ann. Phys. (N.Y.)} {\bf 111}, 111.
\item {[10]}
P. E. T. Jorgensen and R F Werner (1994). {\it Commun. Math. Phys.}
{\bf 164}, 466.
\item {[11]}
R. F. Werner (1995). {\it The Classical Limit of Quantum Theory}
(QUANT-PH 9504016).
\item {[12]}
A. J. MacFarlane (1989). {\it J. Phys. A} {\bf 22}, 4581; 
L. C. Biedenharn (1989). {\it J. Phys. A} {\bf 22}, L873;
Sun C P and Fu H C 1989 {\it J. Phys. A} {\bf 22}, L983.
\item {[13]}
M. Bozejko, B. Kummerer and R. Speicher (1997). 
{\it Commun. Math. Phys.} {\bf 185}, 129. 
\item {[14]}
V. F. Los (1978). {\it Theor. Math. Phys.} {\bf 35}, 113.

\bye